\documentclass[twocolumn,tighten]{aastex701}
\usepackage{isotope}

\graphicspath{{figs/}}
\usepackage{amsmath}
\usepackage{graphicx}
\usepackage{multirow}
\usepackage{float}
\usepackage{array}
\usepackage{makecell}

\begin{document}

\title{Effects of Competing Reactions on $^{59}\mathrm{Cu}$ and $^{63}\mathrm{Ga}$ on Abundance Distributions in Type I X-Ray Bursts}

\author[orcid=0009-0004-5258-9491]{Yinwen Guan}
\affiliation{Shenzhen Key Laboratory of Research and Manufacture of High Purity Germanium Materials and Detectors, Institute for Advanced Study in Nuclear Energy $\&$ Safety, College of Physics and Optoelectronic Engineering, Shenzhen University, Shenzhen 518060, Guangdong, People’s Republic of China}
\email{2500232013@mails.szu.edu.cn}  

\author[orcid=0000-0002-3117-1318]{Ertao Li}
\affiliation{Shenzhen Key Laboratory of Research and Manufacture of High Purity Germanium Materials and Detectors, Institute for Advanced Study in Nuclear Energy $\&$ Safety, College of Physics and Optoelectronic Engineering, Shenzhen University, Shenzhen 518060, Guangdong, People’s Republic of China}
\email{let@szu.edu.cn}
\correspondingauthor{let@szu.edu.cn}

\author[orcid=0000-0001-5206-4661]{Zhihong Li}
\affiliation{China Institute of Atomic Energy, Beijing 102413, People’s Republic of China}
\email{zhli@ciae.ac.cn}

\author[orcid=0000-0001-7149-814X]{Youbao Wang}
\affiliation{China Institute of Atomic Energy, Beijing 102413, People’s Republic of China}
\email{ybwang@ciae.ac.cn} 

\author[orcid=0000-0002-4911-0847]{Bing Guo}
\affiliation{China Institute of Atomic Energy, Beijing 102413, People’s Republic of China}
\email{guobing@ciae.ac.cn}

\author[orcid=0000-0001-8702-431X]{Yunju Li}
\affiliation{China Institute of Atomic Energy, Beijing 102413, People’s Republic of China}
\email{li_yunju@163.com}

\author[orcid=0000-0002-2718-9451]{Jun Su}
\affiliation{School of Physics and Astronomy, Beijing Normal University, Beijing 100875, People’s Republic of China}
\email{sujun@bnu.edu.cn}

\author[orcid=0000-0001-6724-2256]{Shipeng Hu}
\affiliation{Shenzhen Key Laboratory of Research and Manufacture of High Purity Germanium Materials and Detectors, Institute for Advanced Study in Nuclear Energy $\&$ Safety, College of Physics and Optoelectronic Engineering, Shenzhen University, Shenzhen 518060, Guangdong, People’s Republic of China}
\email{husp@szu.edu.cn} 

\author[orcid=0009-0008-9198-5971]{Yu Liu}
\affiliation{Shenzhen Key Laboratory of Research and Manufacture of High Purity Germanium Materials and Detectors, Institute for Advanced Study in Nuclear Energy $\&$ Safety, College of Physics and Optoelectronic Engineering, Shenzhen University, Shenzhen 518060, Guangdong, People’s Republic of China}
\email{2500232005@mails.szu.edu.cn} 

\author[orcid=0009-0003-3896-0269]{Dong Xiang}
\affiliation{Shenzhen Key Laboratory of Research and Manufacture of High Purity Germanium Materials and Detectors, Institute for Advanced Study in Nuclear Energy $\&$ Safety, College of Physics and Optoelectronic Engineering, Shenzhen University, Shenzhen 518060, Guangdong, People’s Republic of China}
\email{2500231015@mails.szu.edu.cn} 

\author[orcid=0009-0009-2232-4317]{Lei Yang}
\affiliation{Shenzhen Key Laboratory of Research and Manufacture of High Purity Germanium Materials and Detectors, Institute for Advanced Study in Nuclear Energy $\&$ Safety, College of Physics and Optoelectronic Engineering, Shenzhen University, Shenzhen 518060, Guangdong, People’s Republic of China}
\email{2310456020@email.szu.edu.cn}

\author[orcid=0000-0003-3233-8260]{Weiping Liu}
\affiliation{Department of Physics, Southern University of Science and Technology, Shenzhen 518055, Guangdong, People’s Republic of China}
\email{wpliu@ciae.ac.cn}

\begin{abstract}
The composition of X-ray burst ashes affects the thermal transport properties of the neutron star crust, as well as heating and cooling through weak interaction processes. Reaction-rate uncertainties can produce secondary peaks separated from the main peak in final abundance distributions. Using STARLIB reaction rates and their temperature-dependent uncertainties in the K04 one-zone post-processing model, we combine four sets of Monte Carlo calculations with XGBoost classification, SHAP analysis, and reaction-flow analysis to investigate the origins of these secondary peaks. We find that the competing $(p,\alpha)$ and $(p,\gamma)$ reaction pairs on $^{59}\mathrm{Cu}$ and $^{63}\mathrm{Ga}$ are primarily responsible for secondary peaks outside the central 68\% confidence interval in all 44 isotopes examined. Fixing both pairs at their median rates reduces the fractions of samples in the secondary-peak regions by 96\%--100\%, while the bounds of the central 68\% interval remain largely unchanged for most mass numbers. The $^{59}\mathrm{Cu}$ pair has the larger effect. Enhanced recycling limits the supply of material to heavier isotopes and accelerates proton consumption, while earlier proton exhaustion further suppresses subsequent proton captures. These changes alter the balance between isotope production and destruction, leading to high- or low-abundance secondary peaks over a broad mass range beyond the Ni--Cu and Zn--Ga cycles. These results identify $^{59}\mathrm{Cu}(p,\gamma)^{60}\mathrm{Zn}$ and the two competing channels on $^{63}\mathrm{Ga}$ as priorities for further nuclear-physics constraints to improve XRB abundance predictions.
\end{abstract}


\section{Introduction} 
Type I X-ray bursts (XRBs) result from thermonuclear runaways in the hydrogen- and helium-rich envelopes of accreting neutron stars. In mixed H/He burning, the $3\alpha$ reaction, the $\alpha p$-process, and the $rp$-process, consisting of rapid proton captures and $\beta^{+}$ decays, drive energy generation and the synthesis of heavier isotopes. Burst ashes are gradually incorporated into the neutron star crust, where their composition affects thermal conductivity and nuclear heating and cooling. Reliable predictions of final abundances are therefore essential for understanding how XRB nucleosynthesis affects neutron star thermal evolution~\citep{Xu2026}.

These processes involve many unstable, neutron-deficient isotopes. Uncertainties in their reaction rates affect reaction paths, energy generation, and the final ash composition. \citet{Parikh2008} studied XRB nucleosynthesis by varying reaction rates individually and by varying all rates simultaneously in Monte Carlo (MC) calculations. \citet{Cyburt2016} subsequently identified key reactions affecting burst light curves and ash compositions using self-consistent burst models. By varying many reaction rates simultaneously, MC calculations propagate their uncertainties to final abundance distributions. STARLIB provides reaction-rate probability distributions and temperature-dependent uncertainties for such calculations~\citep{Longland2012,Sallaska2013}.

In the $rp$-process, the Ni--Cu and Zn--Ga cycles return some material to lighter isotopes. The competing $(p,\alpha)$ and $(p,\gamma)$ reactions on $^{59}\mathrm{Cu}$ and $^{63}\mathrm{Ga}$ regulate recycling and escape from the respective cycles. These cycles affect the supply of material to heavier isotopes and hydrogen burning, thereby changing the final ash composition~\citep{Lam2022c,Xu2026}. Recent direct measurements of $^{59}\mathrm{Cu}(p,\alpha)^{56}\mathrm{Ni}$ have reduced its rate uncertainty, leaving $^{59}\mathrm{Cu}(p,\gamma)^{60}\mathrm{Zn}$ as the dominant remaining uncertainty in the Ni--Cu branching ratio~\citep{LopezSaavedra2026Long}. Quantifying how these competing branches shape abundance distributions also provides clear targets for further nuclear physics measurements.

The large-scale MC calculations of \citet{Wang2026} revealed multipeak final abundance distributions in XRBs. In particular, an increased $^{59}\mathrm{Cu}(p,\alpha)/(p,\gamma)$ rate ratio strengthens Ni--Cu recycling and produces a high-abundance secondary peak of $^{55}\mathrm{Co}$. This finding motivates a systematic investigation of the origins of secondary peaks across the reaction network and the relative contributions of different competing reaction pairs.

To identify which of the 1,711 varied reactions control the secondary peaks, we train XGBoost classifiers with samples labeled according to whether they lie in a secondary-peak region. SHAP analysis is then used to determine the contributions of individual reaction. Using the K04 one-zone post-processing model, we adopt the STARLIB MC calculations of \citet{Wang2026}, in which all selected rates are varied, as the baseline calculation. We perform three additional sets of targeted MC calculations, fixing the $^{59}\mathrm{Cu}$ reaction pair, the $^{63}\mathrm{Ga}$ reaction pair, or both pairs, with 100,000 trials in each set. We quantify the effects of the two reaction pairs, and investigate the underlying processes using XGBoost regression models, reaction-flow analysis, and abundance evolution.

We find that the competing reaction pairs on $^{59}\mathrm{Cu}$ and $^{63}\mathrm{Ga}$ are primarily responsible for the secondary peaks outside the central 68\% confidence interval in all 44 isotopes examined. Fixing both pairs at their STARLIB median rates reduces the fractions of samples in the secondary-peak regions by 96\%--100\%, while the bounds of the central 68\% C.I. remain largely unchanged in most cases. These reaction pairs affect abundance distributions over a broad mass range through changes in recycling, the supply of material to heavier isotopes, and proton exhaustion. For comparison, we also examine bimodal distributions within the central 68\% C.I. and their associated reaction paths.

\section{models and methods} 

\subsection{\label{sec:Models}Thermodynamic Trajectory and Initial Composition}
\begin{figure}[!htbp]
	\centering
	\includegraphics[width=\columnwidth]{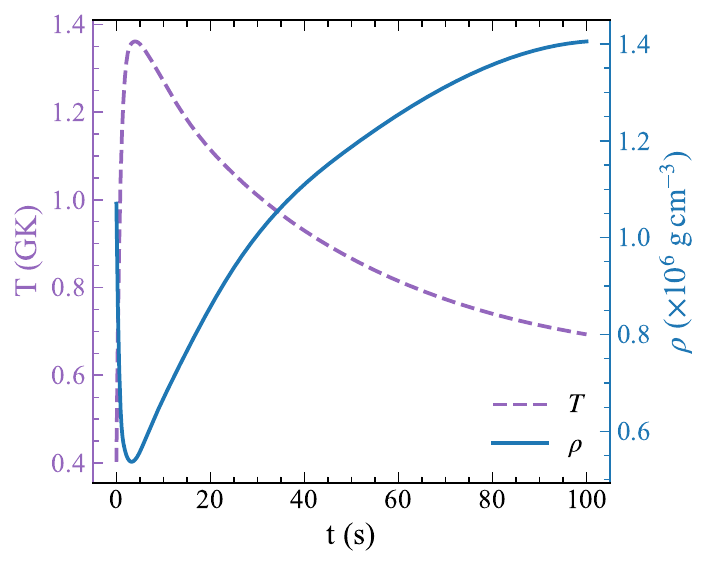}
	\caption{\label{fig:profiles}    
    Time evolution of temperature (in units of $10^9~\mathrm{K}$) and density (in units of $10^6~\mathrm{g/cm^{3}}$) for the XRB Model K04~\citep{Koike2004}.
 }
\end{figure}

All nucleosynthesis calculations followed the model K04~\citep{Koike2004}. The temperature and density profiles are shown in Figure~\ref{fig:profiles}. The peak temperature is \(1.36\,\mathrm{GK}\), and the density ranges from \(0.54\times10^{6}\) to \(1.44\times10^{6}\ \mathrm{g\,cm^{-3}}\) over approximately 100 s. Following \citet{Parikh2008} (based on \citet{Koike2004}), the initial hydrogen, helium, and metal mass fractions were set to \(X=0.73\), \(Y=0.25\), and \(Z=0.02\), respectively, with the metal composition scaled from the solar abundances of \citep{Grevesse1998}. 

\subsection{Nucleosynthesis Calculation Methods}
\label{sec:calculation}
We perform one-zone post-processing calculations with WinNet~\citep{Winteler2012,Reichert2023}, adopting the reaction network and nuclear data settings of \citet{Wang2026}. The network contains 686 nuclides, from hydrogen to $^{136}\mathrm{Xe}$, and more than 8000 reactions. For 1711 forward reactions, including $(p,\gamma)$, $(p,\alpha)$, $(\alpha,p)$, and $(\alpha,\gamma)$, we adopt the median rates and associated uncertainties from STARLIB v6.10~\citep{Sallaska2013}. The corresponding inverse rates are calculated via detailed balance. All other rates are taken from REACLIB v2.2 and remain unchanged~\citep{Cyburt2010}, and the \(3\alpha\) reaction is also excluded from the rate variations~\citep{Parikh2008}.

After the trajectory ends, radioactive species with half-lives shorter than $1\,\mathrm{h}$ are assumed to decay into their respective stable daughter nuclides~\citep{Parikh2008}. All final abundances include this decay treatment. Instead of using absolute isotopic abundances, we consider relative abundances, i.e., abundances from the Monte Carlo trials normalized to baseline abundances. The baseline abundances $X_{\mathrm{base}}$ are calculated with all 1711 forward rates set to their STARLIB median values~\citep{Wang2026}.

\subsection{Monte Carlo Calculation Method}
\label{sec:mc}

\subsubsection{Sampling Method}
\label{sec:sampling}
We follow the STARLIB lognormal sampling procedure~\citep{Longland2012,Sallaska2013}. The reaction rates in STARLIB follow a lognormal probability distribution at each temperature \(T\):
\begin{equation}
f(r)=\frac{1}{\sqrt{2\pi}\,r\sigma}
e^{-\frac{(\ln(r)-\mu)^2}{2\sigma^2}},
\qquad 0<r<\infty ,
\label{eq:lognormal}
\end{equation}
where r is the reaction rate. The lognormal distribution is characterized by a location parameter \(\mu\) and a spread parameter \(\sigma\). Following \citet{Longland2010} and \citet{Sallaska2013}, the median reaction rate and uncertainty factor are given by \(r_{\mathrm{med}}=e^\mu\) and \(f.u.=e^\sigma\), respectively.

For each Monte Carlo trial \(i\), a random variation factor \(p_{ij}\) is assigned to each reaction rate \citep{Longland2012,Sallaska2013,Fields2016}.
\begin{equation}
r_{ij}=e^{\mu}\left(e^{\sigma}\right)^{p_{ij}}
=r_{\mathrm{med}}\left(\mathrm{f.u.}\right)^{p_{ij}},
\label{eq:normal distribution}
\end{equation}
where \(p_{ij} \sim \mathcal{N}(0,1)\) for reaction \(j\).  Following \citet{Longland2012}, we assume \(p_{ij}\) to be constant over the temperature range, which simplifies the sampling procedure while providing a reasonable approximation. When \(p_{ij}=0\), the sampled rate recovers \(r_{\mathrm{med}}\). The temperature dependence of \(f.u.\) is retained in the sampled rate distribution.

\subsubsection{\label{sec:comparative}Targeted reaction-rate fixing}
Based on the multi-peak phenomenon identified in the baseline Monte Carlo calculations of \citet{Wang2026}, we performed additional comparative calculations. The baseline calculation consists of 100,000 Monte Carlo trials, with the rates of 1,711 reactions independently perturbed. In the comparative calculations, the four reactions corresponding to the competing \((p,\alpha)\) and \((p,\gamma)\) channels for \(^{59}\mathrm{Cu}\) and \(^{63}\mathrm{Ga}\) were fixed at their STARLIB median rates, while all other reaction rates were sampled following the procedure described in Section~\ref{sec:sampling}. Each calculation contains 100,000 Monte Carlo trials. In addition, we performed separate calculations by fixing each reaction pair individually to distinguish their respective contributions. It should be noted that, for these four reactions, the uncertainty factor \(f.u.\) adopted in STARLIB is 10.

\subsection{\label{sec:analysis methods}Machine-learning analysis}
In this study, we trained XGBoost binary classifiers to identify reactions associated with the secondary peaks outside the 68\% C.I. in the abundance distributions. Regression models were also used for supplementary abundance analysis~\citep{Chen2016}.

For each Monte Carlo trial \(i\), the input feature vector was constructed from the reaction rate random variation factor \(p_{ij}\):
\begin{equation}
\boldsymbol{p}_i=(p_{i1},p_{i2},\ldots,p_{iM}),
\end{equation}
where \(p_{ij}\) is the reaction-rate sampling parameter defined in Section~\ref{sec:sampling} and \(M\) is the total number of varied reaction rates.

We located the minimum probability density between the main and secondary peaks in the Gaussian-smoothed abundance histogram. Samples with valid, positive abundances were divided into two groups at the corresponding abundance. In addition, the same abundance was used to compare sample fractions before and after fixing the reaction pairs on \(^{59}\mathrm{Cu}\), \(^{63}\mathrm{Ga}\), or both. These samples on the side containing the secondary peak were labeled 1, and the remaining samples were labeled 0. The data were divided into training and test sets in a 75:25 ratio, preserving the class proportions. Classification performance was evaluated on the test set using average precision (AP), precision, and recall~\citep{Saito2015}. Precision and recall were calculated using a classification-score threshold of 0.5. AP summarizes the precision–recall performance across classification thresholds.

For supplementary regression analysis, an independent XGBoost regression model was trained for each final nuclide to model the relationship between the reaction rate variations and its final relative abundance. The target variable was the final abundance relative to the baseline abundance, expressed as
\begin{equation}
y_i=\log_{10}\left(\frac{Y_i}{Y_0}\right),
\end{equation}
where \(Y_i\) is the final abundance of the target nuclide in trial i and \(Y_0\) is its corresponding baseline abundance. Monte Carlo samples were randomly divided into training and test sets with a ratio of 75:25. The model was evaluated on the independent test set using the root-mean-square error (RMSE) between the predicted and MC logarithmic relative abundances.

To investigate the contributions of individual reaction parameters to the model predictions, we further employed Shapley Additive exPlanations (SHAP) based on the TreeSHAP framework~\citep{Lundberg2019,Lundberg2020}. SHAP decomposes each model prediction into additive contributions from the input features:
\begin{equation}
f(\boldsymbol{p}_i) = \phi_0 + \sum_{j=1}^{M} \phi_{ij}.
\label{eq:shap}
\end{equation}
where \(\phi_0\) is the model’s reference prediction and \(\phi_{ij}\) is the contribution of the sampling parameter for reaction \(j\) to the model prediction for sample \(i\). For classification, SHAP values are expressed in log-odds, with positive values favoring label 1. For regression, SHAP values show how each reaction contributes to the predicted \(\log_{10}(Y/Y_0)\). In addition, reactions were ranked by their mean absolute SHAP values, and beeswarm plots~\citep{Lundberg2020} were used to show the magnitudes and directions of these contributions.

Specifically, in this work, XGBoost and SHAP are primarily used as tools for reaction screening. The reactions identified by these analyses were further examined using targeted MC calculations and reaction-flow analysis to check their effects on the abundance distributions and investigate the physical processes involved.

\section{Results and Discussion} 
In this work, We analyze the final abundances of nuclide after radioactive decay of the network products. We use the Monte Carlo results obtained using STARLIB from \citet{Wang2026} as the baseline. For simulations using STARLIB, abundance uncertainties are estimated at the 68\% C.I., consistent with the confidence interval of \(f.u.\) ~\citep{Psaltis2025}.

We examined an isotope’s abundance distribution only if at least 500 MC samples had valid, positive abundances. For classification, both the secondary peak and the location of minimum probability density between the two peaks had to lie outside the central 68\% C.I.. Secondary peaks on either side of the main peak were included. Each class contained at least 100 samples, and secondary-peak samples made up at least 0.1\% of the positive-abundance samples.

\subsection{\label{sec:classification}Reactions Associated with Secondary Peaks}
\begin{figure}[!htbp]
\centering
\includegraphics[width=\columnwidth]{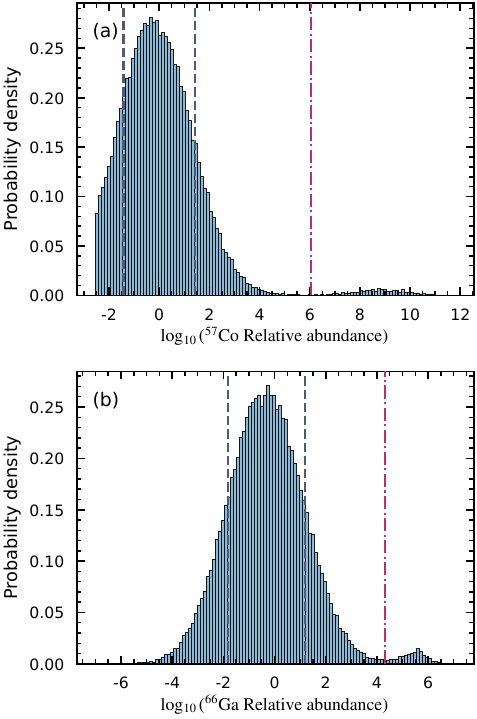}\\
\caption{\label{fig:co57 and ga66}Histograms of the logarithmic relative abundances for \(^{57}\mathrm{Co}\) (a) and \(^{66}\mathrm{Ga}\) (b) in the baseline MC calculations for Model K04. The magenta dash-dotted line in each panel indicates the location of minimum probability density between the two peaks. The dark blue dashed lines indicate the central 68\% C.I..}
\end{figure}

The abundance distributions from the baseline MC calculations exhibit secondary peaks outside the central 68\% C.I. for several isotopes. These peaks occur on either the low- or high-abundance side of the main peak. Figure~\ref{fig:co57 and ga66} shows the distributions of \(^{57}\mathrm{Co}\) and \(^{66}\mathrm{Ga}\) as representative examples. Both exhibit a smaller peak at higher abundances. The location of minimum probability density between the two peaks lies above the upper bound of the central 68\% C.I.

Classifiers were trained for 44 selected isotopes, including 24 with high-abundance secondary peaks and 20 with low-abundance secondary peaks. The median test-set AP, precision, and recall were 0.917, 0.804, and 0.876, respectively. For \(^{57}\mathrm{Co}\) and \(^{66}\mathrm{Ga}\), the AP values were 0.923 and 0.926. When the rankings were calculated over all valid MC samples and separately over secondary-peak samples, we found that the \(^{59}\mathrm{Cu}(p,\alpha)^{56}\mathrm{Ni}\) and \(^{59}\mathrm{Cu}(p,\gamma)^{60}\mathrm{Zn}\) reactions occupied the first two SHAP ranks in all 44 classifiers, followed by \(^{63}\mathrm{Ga}(p,\alpha)^{60}\mathrm{Zn}\) and \(^{63}\mathrm{Ga}(p,\gamma)^{64}\mathrm{Ge}\). Notably, the SHAP rankings indicate that the competing \(^{67}\mathrm{As}(p,\alpha)^{64}\mathrm{Ge}\) and \(^{67}\mathrm{As}(p,\gamma)^{68}\mathrm{Se}\) reactions also contribute to the secondary peaks, although less strongly than the \(^{59}\mathrm{Cu}\) and \(^{63}\mathrm{Ga}\) reaction pairs. Complete evaluation results are provided in Appendix\ref{app:classification}.

Figure~\ref{fig:shap_co57_ga66} shows the classification SHAP beeswarm plots for \(^{57}\mathrm{Co}\) and \(^{66}\mathrm{Ga}\), illustrating which reaction-rate variations favor assignment to the secondary-peak region. For both isotopes, higher rates of the \((p,\alpha)\) reactions on \(^{59}\mathrm{Cu}\) and \(^{63}\mathrm{Ga}\) favor classification as secondary-peak samples. Lower rates of the corresponding \((p,\gamma)\) reactions also favor this classification. These trends are seen in both the full and secondary-peak sample sets.

\begin{figure*}[!htbp]
\centering
\includegraphics[width=\textwidth]{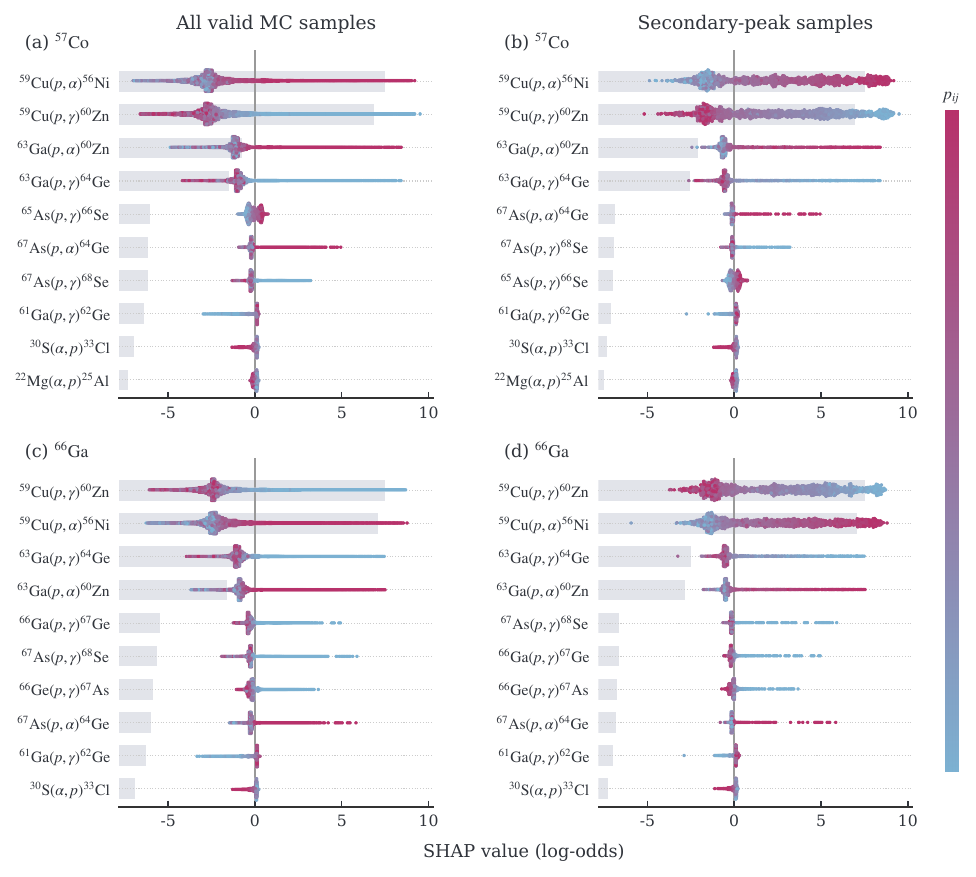}\\
\caption{\label{fig:shap_co57_ga66}SHAP beeswarm plots for the \(^{57}\mathrm{Co}\) (a,b) and \(^{66}\mathrm{Ga}\) (c,d) classifiers. The left column shows all valid MC samples, and the right column shows only secondary-peak samples. Each panel shows the ten reactions with the largest mean absolute SHAP values for the corresponding sample set, in descending order. Each point represents one MC sample. SHAP values are in log-odds, with positive values favoring label 1 (secondary-peak samples). Colors indicate the reaction-rate sampling parameter \(p_{ij}\), from blue (low) to red (high).}
\end{figure*}

\subsection{\label{sec:fix reactions}Effects of Fixing Reaction Rates}
To examine how these reaction-rate uncertainties affect the secondary peaks, we performed three additional sets of MC calculations. We fixed the rates of \(^{59}\mathrm{Cu}(p,\alpha)^{56}\mathrm{Ni}\) and \(^{59}\mathrm{Cu}(p,\gamma)^{60}\mathrm{Zn}\) , the rates of \(^{63}\mathrm{Ga}(p,\alpha)^{60}\mathrm{Zn}\) and \(^{63}\mathrm{Ga}(p,\gamma)^{64}\mathrm{Ge}\), or all four rates at their STARLIB median values. We then compared the fractions of MC samples in the secondary-peak regions.

Simultaneously fixing the \(^{59}\mathrm{Cu}\) and \(^{63}\mathrm{Ga}\) reaction pairs reduced the secondary-peak sample fractions for all 44 selected isotopes, with relative reductions of 96.0-100\%. Fixing only the \(^{59}\mathrm{Cu}\) reaction pair produced a larger reduction than fixing only the \(^{63}\mathrm{Ga}\) reaction pair in every case. The median relative reductions were 83.9\% and 25.5\%, respectively. The results for individual isotopes are provided in Appendix~\ref{app:fractions}.

Figure~\ref{fig:extra counts} compares the numbers and fractions of MC samples in the high-abundance secondary-peak regions of \(^{57}\mathrm{Co}\) and \(^{66}\mathrm{Ga}\) for the four sets of calculations. When both pairs were fixed, the fraction of samples in the high-abundance secondary-peak region decreased from 1.489\% to 0.027\% for \(^{57}\mathrm{Co}\), and from 1.512\% to 0.060\% for \(^{66}\mathrm{Ga}\). Fixing only the \(^{59}\mathrm{Cu}\) reaction pair reduced these fractions to 0.279\% and 0.313\%, whereas fixing only the \(^{63}\mathrm{Ga}\) reaction pair reduced them to 1.080\% and 1.118\%, respectively. The corresponding abundance distributions for the four sets of calculations are shown in Appendix~\ref{app:fractions}.

\begin{figure}[!htbp]
\centering
\includegraphics[width=\columnwidth]{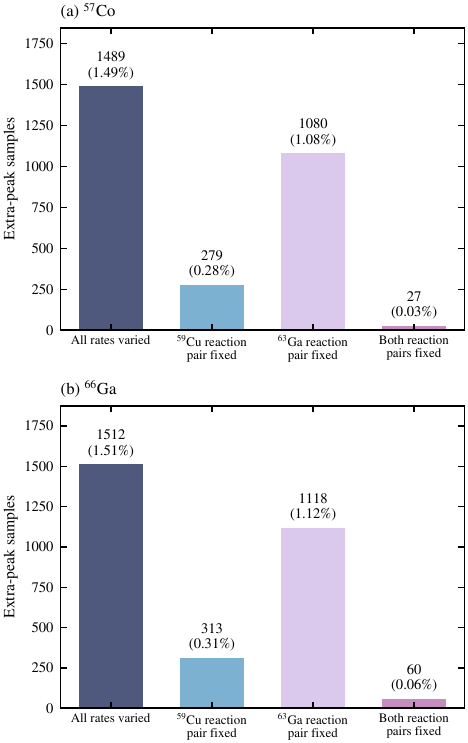}\\
\caption{\label{fig:extra counts}Numbers of MC samples in the high-abundance secondary-peak regions of (a) \(^{57}\mathrm{Co}\) and (b) \(^{66}\mathrm{Ga}\). The four sets of calculations include the baseline case and cases with the \(^{59}\mathrm{Cu}\) reaction pair, the \(^{63}\mathrm{Ga}\) reaction pair, or both pairs fixed at their STARLIB median rates. For each isotope, secondary-peak samples in all four sets are counted using the location of minimum probability density between the two peaks in the abundance distribution from the baseline MC calculations. Labels above the bars give the sample counts, with the corresponding percentages of all MC samples in parentheses.}
\end{figure}

\subsection{\label{sec:flow}Origin of the Secondary Peaks}
\begin{figure*}[!htbp]
\centering
\includegraphics[
        width=0.75\textwidth,
        height=0.75\textheight
    ]{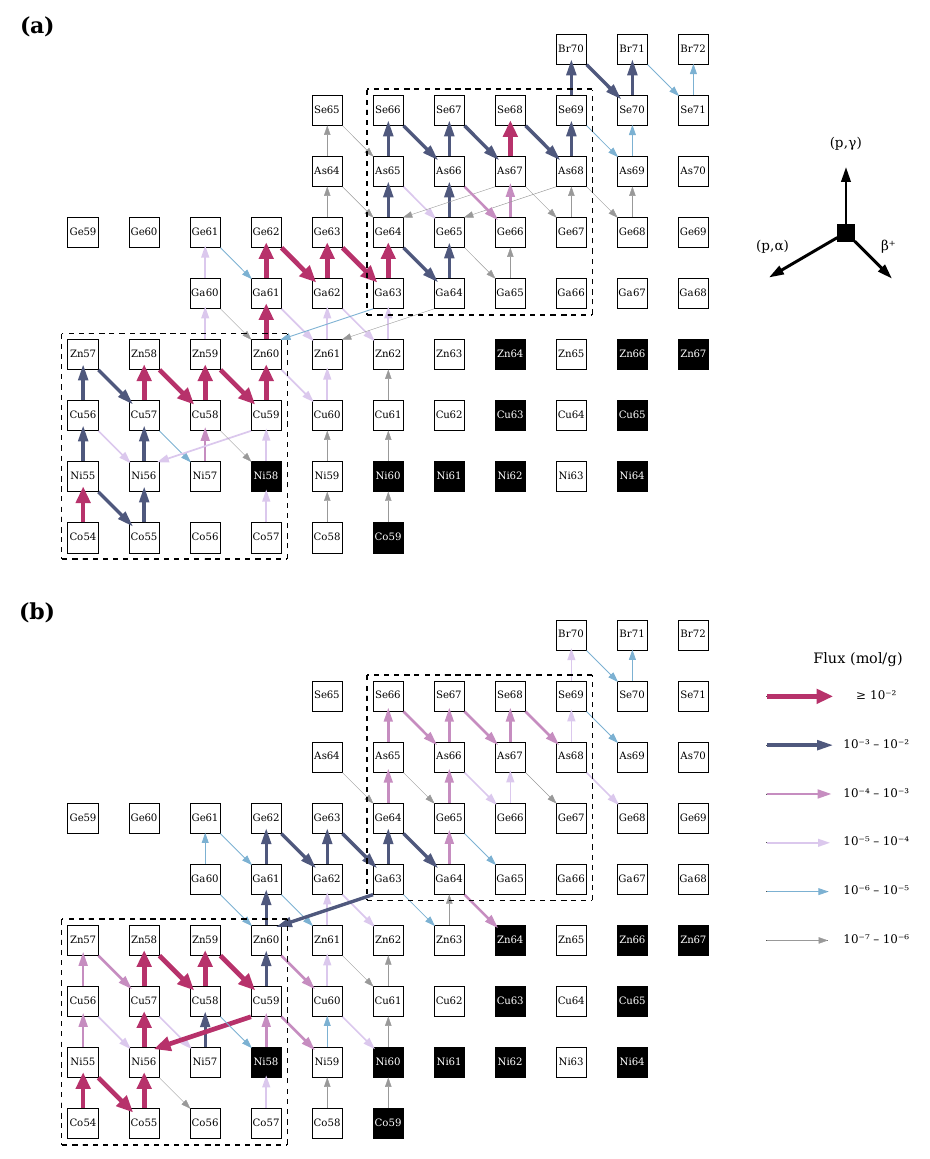}\\
\caption{\label{fig:cycle flow}Integrated net reaction fluxes at the end of the K04 trajectory for (a) the median-rate calculation and (b) the selected MC sample. Stable nuclides are marked by black squares. Arrow colors and widths indicate flux magnitudes, with the scale shown on the right.}
\end{figure*}

To explain the secondary peaks discussed in Section~\ref{sec:fix reactions}, we compare the median-rate calculation (STARLIB) with a representative MC sample. This sample was selected from the baseline MC calculations and belongs to the secondary-peak regions of both \(^{57}\mathrm{Co}) and (^{66}\mathrm{Ga}\). Figure~\ref{fig:cycle flow} compares the corresponding integrated net fluxes.

\begin{figure}[!htbp]
\centering
\includegraphics[width=\columnwidth]{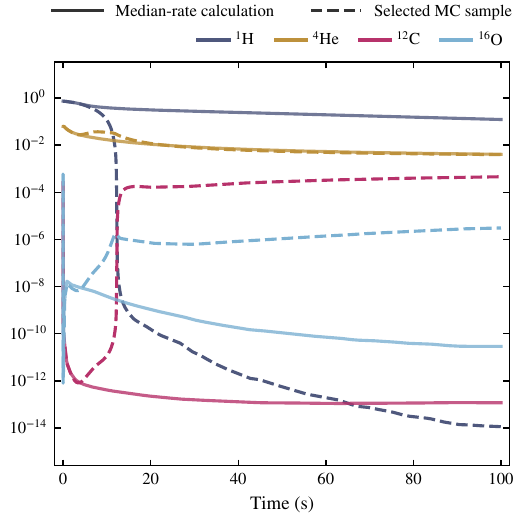}\\
\caption{\label{fig:abundance evolution}Abundance evolution of \(^{1}\mathrm{H}\), \(^{4}\mathrm{He}\), \(^{12}\mathrm{C}\), and \(^{16}\mathrm{O}\) along the K04 trajectory. Solid and dashed lines show the median-rate calculation and the selected MC sample, respectively. Colors distinguish the isotopes, as indicated in the legend.}
\end{figure}

In the selected sample, the \((p,\alpha)/(p,\gamma)\) reaction-rate ratios for \(^{59}\mathrm{Cu}\) and \(^{63}\mathrm{Ga}\) are approximately \(1.0\times10^{4}\) and \(4.2\times10^{3}\) times their corresponding median-rate ratios, respectively. Figure~\ref{fig:cycle flow} shows stronger flows through \(^{59}\mathrm{Cu}(p,\alpha)^{56}\mathrm{Ni}\) and \(^{63}\mathrm{Ga}(p,\alpha)^{60}\mathrm{Zn}\), together with weaker flows through the competing \((p,\gamma)\) channels and the Ge--Se region. Enhanced recycling therefore reduces the transport of material toward heavier nuclei while continuing to consume protons~\citep{Cyburt2016,Lam2022c,Lam2022b}. The abundance evolution in Figure~\ref{fig:abundance evolution} also shows earlier proton exhaustion in the selected sample, shortening the period during which proton captures can destroy the nuclei produced. Reduced transport toward heavier nuclei therefore does not imply lower final abundances and the outcome depends on the balance between production and destruction. \citet{Merz2021} likewise emphasized the role of proton-capture destruction in determining their survival for odd-\(A\) mass chains (including \(A=57\)). The regression SHAP results provide further support for this interpretation, with higher \(^{57}\mathrm{Co}(p,\gamma)\) and \(^{66}\mathrm{Ga}(p,\gamma)\) rates generally contributing to lower predicted abundances of the corresponding isotopes (Figure~\ref{fig:regression_shap_co57ga66}). Together, these results connect the high-abundance secondary peaks of \(^{57}\mathrm{Co}\) and \(^{66}\mathrm{Ga}\) to enhanced recycling and reduced proton-capture destruction following early proton exhaustion. When the reduction in destruction exceeds the
reduction in production, these isotopes can still accumulate and form high-abundance secondary peaks, even if less material reaches the corresponding mass region.

In addition, late helium burning after hydrogen consumption enhances the abundances of \(\alpha\)-chain isotopes, including \(^{12}\mathrm{C}\) and \(^{16}\mathrm{O}\) ~\citep{Cyburt2016}, as shown in Figure~\ref{fig:abundance evolution}. In the selected sample, the proton abundance drops sharply at approximately $10$--$15\,\mathrm{s}$ and reaches about $10^{-14}\,\mathrm{mol/g}$ at the end of the trajectory, whereas the median-rate calculation retains an abundance of order $10^{-1}\,\mathrm{mol/g}$. As the protons are depleted, helium continues to be consumed and $^{12}\mathrm{C}$ and $^{16}\mathrm{O}$ recover from their early minima. Their final abundances are much higher than in the median-rate calculation. That is to say, helium burning continues to produce $^{12}\mathrm{C}$ and $^{16}\mathrm{O}$, while proton depletion reduces their destruction by proton captures.

For heavier isotopes, reduced supply beyond the cycles combines with proton exhaustion to suppress further synthesis. Material must still pass through waiting-point regions such as \(^{64}\mathrm{Ge}\), where the competition between \(\beta\) decay and sequential proton captures controls further processing \citep{Lam2016}. This gives the opposite abundance response: low-abundance secondary peaks and long tails, in contrast to the high-abundance secondary peaks discussed above. Changes in reaction flow and proton abundance can therefore help some isotopes survive while reducing the production of others.

\begin{figure}[!htbp]
\centering
\includegraphics[width=\columnwidth]{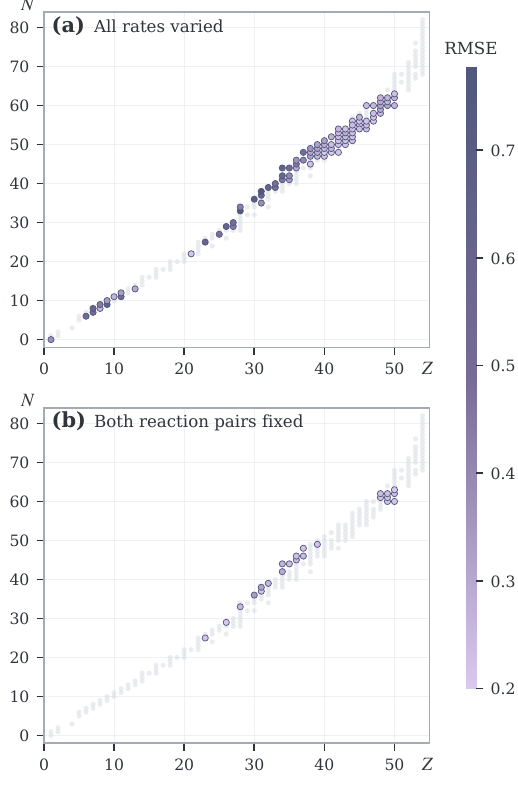}\\
\caption{\label{fig:h_rmse}RMSE for (a) all rates varied and (b) the \(^{59}\mathrm{Cu}\) and \(^{63}\mathrm{Ga}\) reaction pairs fixed. Gray circles show the isotopes included in the network. Purple circles show isotopes with RMSE \(>0.2\), with colors indicating the RMSE values in dex.}
\end{figure}

The influence of the \(^{59}\mathrm{Cu}\) and \(^{63}\mathrm{Ga}\) reaction pairs extends well beyond the Ni--Cu and Zn--Ga cycles. With regression models trained separately for the baseline and reaction pairs fixed MC calculations, fixing both pairs reduces the number of isotopes with RMSE \(>0.2\,\mathrm{dex}\) from 95 to 23 and lowers the RMSE values for most isotopes (Figure~\ref{fig:h_rmse}). In the baseline test sets, secondary-peak samples constitute only \(1.67\%\) and \(1.54\%\) of the \(^{57}\mathrm{Co}\) and \(^{66}\mathrm{Ga}\) samples, respectively, but account for approximately \(67\%\) and \(66\%\) of the total squared prediction error. Thus, a small fraction of samples dominates the regression error for these two isotopes, and suppressing these secondary peaks and extreme tails can reduce their contribution to the regression error. Together with the abundance and flow analyses above, this widespread improvement shows that changes in recycling and proton abundance allow these competing reaction pairs to affect isotope destruction and survival across the network. 
In addition, the \(^{67}\mathrm{As}\) reaction pair noted in Section~\ref{sec:classification} may also contribute to the remaining extreme samples, as its \((p,\alpha)\) branch returns material to \(^{64}\mathrm{Ge}\) through the GeAs I cycle~\citep{Lam2022b}.

\subsection{Impact on Abundance Uncertainties}
\label{sec:68 C.I.}
\begin{figure}[htbp]
\centering
\IfFileExists{central68_comparison.pdf}{%
  \includegraphics[width=\columnwidth]{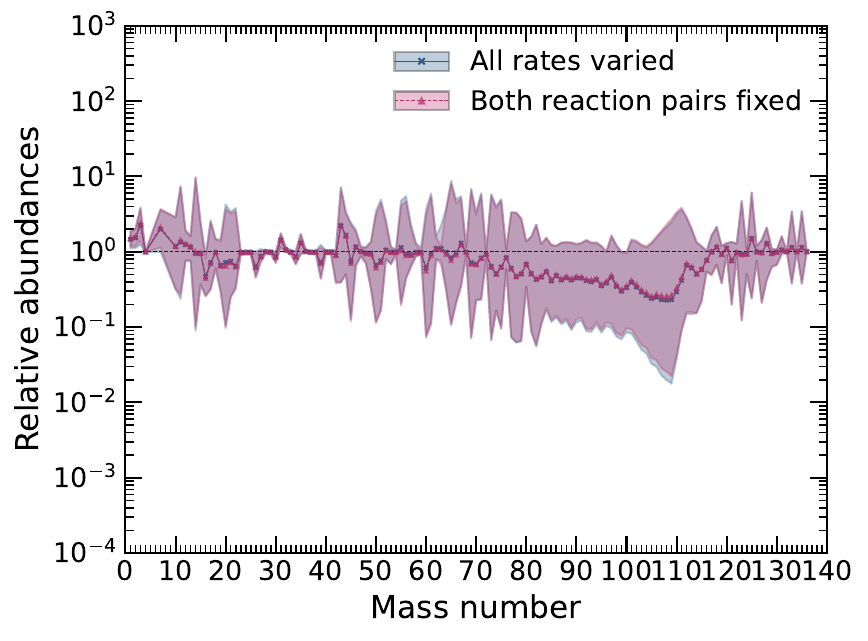}%
}{%
  \fbox{\parbox{0.9\columnwidth}{\centering Upload the abundance-interval comparison figure as central68\_comparison.pdf.}}%
}
\caption{Final abundances summed over mass number for the baseline MC calculations and the calculations with the $^{59}\mathrm{Cu}$ and $^{63}\mathrm{Ga}$ reaction pairs fixed at their median rates. Shaded bands show the abundance uncertainties, which correspond to the 68\% C.I.; curves show the medians.}
\label{fig:central68}
\end{figure}

Figure~\ref{fig:central68} compares the 68\% C.I. of the final abundances before and after fixing the $^{59}\mathrm{Cu}$ and $^{63}\mathrm{Ga}$ reaction pairs. The upper and lower limits remain similar. For 90\% of the 122 mass numbers compared, both limits remain within a factor of 1.15 of the baseline MC calculation values. An exception is \(^{63}\mathrm{Cu}\), for which the upper limit of the 68\% C.I. decreases by about 46\%. The highest-ranked reaction in its regression SHAP analysis, \(^{63}\mathrm{Ga}(p,\gamma)^{64}\mathrm{Ge}\), is one of the fixed reactions.

For $^{57}\mathrm{Co}$ and $^{66}\mathrm{Ga}$, we compare samples within the 68\% C.I. with those in the secondary peaks, and examine the actual $(p,\alpha)/(p,\gamma)$ reaction-rate ratios for $^{59}\mathrm{Cu}$ and $^{63}\mathrm{Ga}$. At 1.25~GK, the median ratios within the 68\% C.I. are approximately $8.6\times10^{-4}$ and $2.1\times10^{-4}$, respectively, indicating that the $(p,\gamma)$ branch generally dominates and recycling is weak. In contrast, the median $^{59}\mathrm{Cu}$ ratio reaches approximately 2.5-2.6 in the secondary-peak samples. More than 92\% of these samples have at least one ratio of 1 or larger, showing that the $(p,\alpha)$ branch can compete with even dominate the $(p,\gamma)$ branch.

The secondary peaks contain only about 1.5\% of all samples. Fixing the reaction pairs can reduce the secondary peaks, but has little effect on the upper and lower limits of the 68\% C.I.. The abundance uncertainties (68\% C.I.) remain large because the other reaction rates are still varied.

The contrast in branching ratios between the central and secondary-peak samples raises a direct question about whether current experimental constraints still allow strong recycling. Earlier sensitivity studies illustrate the effects of enhanced recycling on burst light curves. \citet{Cyburt2016} varied individual reaction rates by a factor of 100 up and down in a single-zone burst model. They found that a stronger NiCu cycle limits the synthesis of heavier nuclei and produces large changes in the burst light curve. \citet{Bhathi2026} increased their new recommended \(^{59}\mathrm{Cu}(p,\alpha)\) rate by a factor of two in the one-zone GS~1826$-$24 model. They found little effect on the light curve, whereas a hundredfold increase produced a large change. That is to say, a hundredfold enhancement above the NON-SMOKER reference rate lies far outside the new experimental bounds. These constraints restrict the enhancement of NiCu recycling through an increased \(^{59}\mathrm{Cu}(p,\alpha)\) rate.

However, enhanced recycling can also arise from a strongly reduced $^{59}\mathrm{Cu}(p,\gamma)$ rate. The recommended \(^{59}\mathrm{Cu}(p,\alpha)\) rate from\citet{LopezSaavedra2026Short,LopezSaavedra2026Long} is nearly identical to STARLIB at $T\gtrsim1$~GK, but its uncertainty factor is about 1.58 at 1.0~GK and 1.47 at 1.5~GK. Compared with the factor of 10 adopted in our MC calculations, the smaller uncertainty factor greatly reduces the probability of extreme combinations, even if the recommended rate remains similar. But, the $^{63}\mathrm{Ga}$ reaction pair still remains unconstrained, and with $(p,\alpha)$ better constrained, the $^{59}\mathrm{Cu}(p,\gamma)$ rate becomes the main source of uncertainty in the Cu branching ratio \citep{LopezSaavedra2026Long}.

\begin{figure}[!htbp]
\centering
\includegraphics[width=\columnwidth]{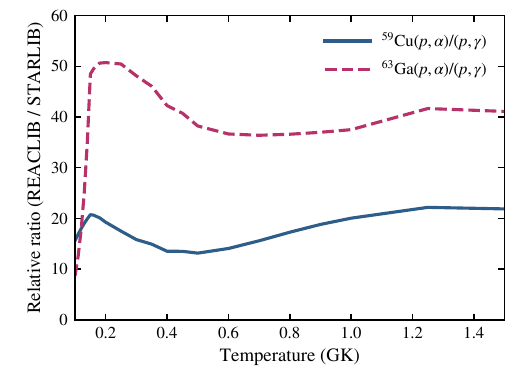}\\
\caption{\label{fig:ratio_compare}REACLIB v2.2 $(p,\alpha)/(p,\gamma)$ reaction-rate ratios relative to the corresponding STARLIB v6.10 ratios for $^{59}\mathrm{Cu}$ (solid) and $^{63}\mathrm{Ga}$ (dashed).}
\end{figure}

In addition, the probability of strong recycling depends not only on the rate uncertainties but also on the adopted median rates. The $^{55}\mathrm{Co}$ secondary peak is less pronounced in our STARLIB calculations than in the REACLIB calculations of \citet{Wang2026}. Both calculations adopt $f.u.=10$ for the relevant reaction pairs. At the K04 peak temperature, the $(p,\alpha)/(p,\gamma)$ reaction-rate ratios for $^{59}\mathrm{Cu}$ and $^{63}\mathrm{Ga}$ are about 20 and 40 times larger in REACLIB v2.2 than in STARLIB v6.10 (Figure~\ref{fig:ratio_compare}). The lower STARLIB ratios favor weaker recycling, consistent with the less pronounced secondary peak.

Differences in nuclear inputs, burning stages, and thermal feedback across studies lead to different effects of reaction-rate variations. Different approaches to extrapolating experimental data give different reaction-rate evaluations \citep{Bhathi2026,LopezSaavedra2026Long}. The same rate variation can change the light curve without affecting the final abundance of $A=59$ nuclei because they are at different burning stages \citep{Merz2021}. Our K04 trajectory excludes the feedback from nuclear heating present in self-consistent burst models \citep{Cyburt2016}. Overall, the experimentally constrained $^{59}\mathrm{Cu}(p,\alpha)$ rates do not support strong NiCu recycling, consistent with the limited impact of the competing reaction pairs on abundances within the 68\% C.I. in our MC calculations. Strong recycling substantially changes the burst light curve and appears in our MC calculations as secondary peaks outside the 68\% C.I.

\subsection{Bimodality within the 68\% C.I.}
\label{sec:bimodal_distributions}
\begin{figure}[!htbp]
\centering
\includegraphics[width=\columnwidth]{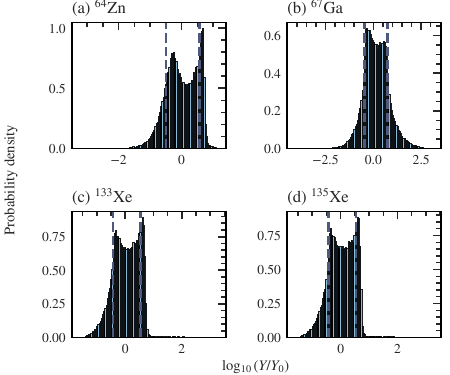}\\
\caption{\label{fig:four_PDF}Histograms of the logarithmic relative abundances for \(^{64}\mathrm{Zn}\) (a), \(^{67}\mathrm{Ga}\) (b), \(^{133}\mathrm{Xe}\) (c), and \(^{135}\mathrm{Xe}\) (d) in Model K04 for the all-varied (baseline) MC calculation. The dashed lines indicate the 68\% C.I. in each panel.}
\end{figure}

\begin{figure}[!htbp]
\centering
\includegraphics[width=\columnwidth]{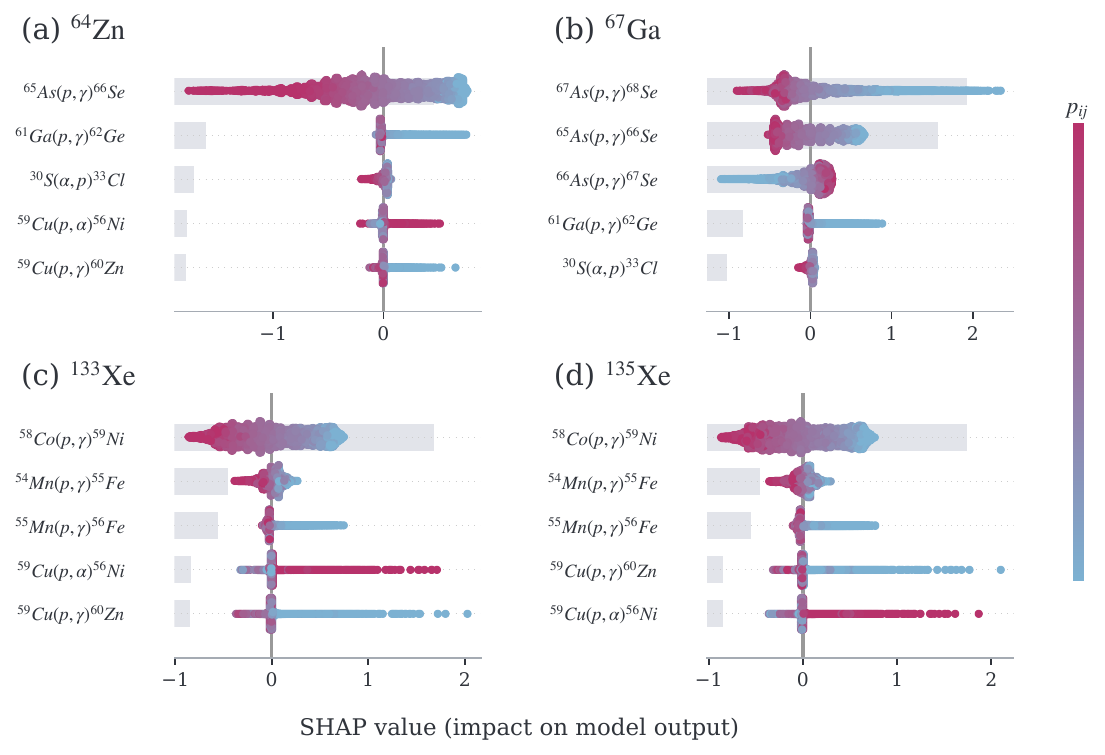}\\
\caption{\label{fig:four_b}Regression SHAP results for \(^{64}\mathrm{Zn}\) (a), \(^{67}\mathrm{Ga}\) (b), \(^{133}\mathrm{Xe}\) (c), and \(^{135}\mathrm{Xe}\) (d) in Model K04. The five reactions with the largest mean absolute SHAP values are shown for each isotope. Gray bars indicate mean absolute SHAP values, while point colors indicate the sampled rate parameters \(p_{ij}\). Positive (negative) SHAP values increase (decrease) the predicted logarithmic relative abundance.}
\end{figure}

Figure~\ref{fig:four_PDF} shows the bimodal abundance
distributions of $^{64}$Zn, $^{67}$Ga, $^{133}$Xe, and $^{135}$Xe.
In these cases, both peaks lie within the 68\% C.I., in contrast to
the secondary peaks outside the 68\% C.I. discussed above, which
arise from competing reaction pairs. We use regression SHAP to
identify the key reactions affecting these abundances
(Figure~\ref{fig:four_b}).

For $^{64}$Zn, regression SHAP indicates that the bimodal distribution is mainly caused by the single reaction \(^{65}\mathrm{As}(p,\gamma)^{66}\mathrm{Se}\), consistent with ~\citet{Wang2026}. This reaction changes the paths around the \(^{64}\mathrm{Ge}\) waiting point and hence the abundances of \(^{64}\mathrm{Ge}\) and \(^{64}\mathrm{Ga}\) remaining to subsequently decay into \(^{64}\mathrm{Zn}\).

The bimodal distributions of $^{133}$Xe and $^{135}$Xe have similar
shapes, and regression SHAP identifies
$^{58}\mathrm{Co}(p,\gamma)^{59}\mathrm{Ni}$ as the most influential
reaction for both isotopes. Higher sampled rates generally
correspond to lower predicted abundances,
with the opposite trend for lower rates.
These consistent responses suggest that this single reaction
predominantly shapes the bimodal distributions of both Xe isotopes.

For $^{67}$Ga, the bimodal distribution reflects the coupled effects
of $^{67}\mathrm{As}(p,\gamma)^{68}\mathrm{Se}$,
$^{65}\mathrm{As}(p,\gamma)^{66}\mathrm{Se}$, and
$^{66}\mathrm{As}(p,\gamma)^{67}\mathrm{Se}$.
Higher sampled rates for the first two reactions generally
contribute to lower predicted abundances, whereas higher rates for
the third contribute to higher abundances.
These reactions regulate the paths leading to the $A=67$ chain and
the subsequent proton-capture flow toward $A=68$.

\begin{figure}[!htbp]
\centering
\includegraphics[width=\columnwidth]{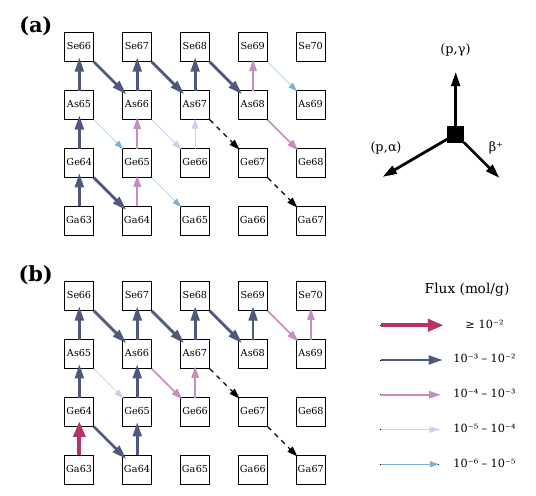}\\
\caption{\label{fig:ga67_flow}Integrated net fluxes in Model K04 at the end of the trajectory for the all-varied MC calculation. (a) A representative Monte Carlo trial corresponding to the peak at low \(^{67}\mathrm{Ga}\) abundance. (b) A representative Monte Carlo trial corresponding to the peak at high \(^{67}\mathrm{Ga}\) abundance. The two trials follow different dominant reaction paths. Flux strengths are indicated by arrow sizes and colors, with the corresponding legend shown on the right. Black dashed arrows indicate the decay sequence leading to \(^{67}\mathrm{Ga}\).}
\end{figure}

To investigate the origin of the $^{67}$Ga bimodality, we compare
the integrated net fluxes of representative samples from the two
peaks (Figure~\ref{fig:ga67_flow}).
In the low-abundance sample, the flow predominantly follows
$^{64}\mathrm{Ge}(p,\gamma)^{65}\mathrm{As}(p,\gamma)^{66}\mathrm{Se}$
and continues through the As-Se region toward $^{68}$Se.
In the high-abundance sample, the branch
$^{64}\mathrm{Ga}(p,\gamma)^{65}\mathrm{Ge}(p,\gamma)^{66}\mathrm{As}$
is stronger, as is
$^{66}\mathrm{Ge}(p,\gamma)^{67}\mathrm{As}$, increasing the
contribution of the lower-$Z$ path to $^{67}$As production.
At $^{67}$As, proton capture carries material to $^{68}$Se, whereas
$\beta^+$ decay leads to $^{67}$Ga through $^{67}$Ge.
Thus, the final $^{67}$Ga abundance is lower when more material
undergoes further proton capture and higher when more material
remains in the $A=67$ decay chain.

\section{Conclusion} 
In this work, we investigated how the competing \((p,\alpha)\) and \((p,\gamma)\) reaction pairs on \(^{59}\mathrm{Cu}\) and \(^{63}\mathrm{Ga}\) shape the final abundance distributions under large-scale propagation of reaction-rate uncertainties in Model K04. To this end, we combined four sets of 100,000-trial MC calculations with XGBoost classification and regression, SHAP analysis, targeted rate fixing, and reaction-flow analysis. For all 44 nuclides with secondary peaks outside the 68\% C.I., the two \(^{59}\mathrm{Cu}\) reactions occupied the top two SHAP ranks in the classification models, followed by the two \(^{63}\mathrm{Ga}\) reaction pairs. Targeted MC calculations further quantified the influence of these reaction pairs on the occurrence of secondary-peak samples.

Fixing both reaction pairs at their STARLIB median rates reduced the fractions of samples in the secondary-peak regions by 96\%--100\% for all 44 nuclides. Fixing only the \(^{59}\mathrm{Cu}\) or \(^{63}\mathrm{Ga}\) pair yields median relative reductions of 83.9\% and 25.5\%, respectively, demonstrating the stronger influence of the \(^{59}\mathrm{Cu}\) pair. Meanwhile, for 90\% of the 122 mass numbers examined, both bounds of the 68\% C.I. remained within a factor of 1.15 of their baseline values. The number of nuclides with regression RMSE (>0.2) dex also decreased from 95 to 23. These results show that the competing reaction pairs strongly influence the secondary peaks and tails of the abundance distributions, while the 68\% C.I. remains broadly stable.

Reaction-flow and abundance-evolution analyses show that different rate combinations produce different degrees of recycling, heavy-nucleus transport, and proton exhaustion. Enhanced recycling limits the supply of material to heavier isotopes, while earlier proton exhaustion suppresses subsequent proton captures, changing the balance between isotope production and destruction. These effects extend throughout the reaction network. Final abundances do not show a simple, linear response to these changes. Differences in proton-capture duration and transport through waiting points can lead some samples to abundances far above or below those of most samples, thereby forming secondary peaks and long tails. Bimodal distributions within the 68\% C.I., however, exhibit different reaction sensitivities and transport pathways.

These results identify \(^{59}\mathrm{Cu}(p,\gamma)\) and the two competing channels on \(^{63}\mathrm{Ga}\) as priorities for further nuclear-physics constraints. Joint constraints on the competing branches will advance quantitative assessments of the probability of strong recycling and the resulting final abundances, improving abundance predictions for X-ray burst. Future MC calculations can further incorporate correlations among reaction-rate uncertainties and thermal feedback in self-consistent burst models to assess their effects on the probability of extreme rate ratios and, consequently, on the final abundance distributions.

\begin{acknowledgments}
The authors thank the reviewer for her/his valuable comments on
the submitted manuscript. The authors thank H. Schatz for providing
the XRB trajectories used in this study. The authors thank Qing Wang for helpful discussions and suggestions. The authors  acknowledge support from the National Natural Science Foundation of China (No. 12175152), LingChuang Research Project of the China National Nuclear Corporation (Nos. 2025-084, 2024-065, 2024-068).
\end{acknowledgments}

\software{\texttt{IPython}~\citep{ipython}, \texttt{Jupyter}~\citep{jupyter}, \texttt{matplotlib}~\citep{matplotlib}, \texttt{numpy}~\citep{numpy}, \texttt{scipy}~\citep{2020SciPy-NMeth}, \texttt{scikit-learn}~\citep{scikit-learn}, \texttt{seaborn}~\citep{Waskom2021}, \texttt{WinNet}~\citep{Winteler2012,Reichert2023}}

\clearpage
\raggedbottom
\appendix
\nopagebreak[4]

\noindent
\begin{minipage}{\textwidth}

\section{Classification Performance}
\label{app:classification}

\begin{figure}[H]
\centering
\includegraphics[
    width=\columnwidth,
    height=0.80\textheight,
    keepaspectratio
]{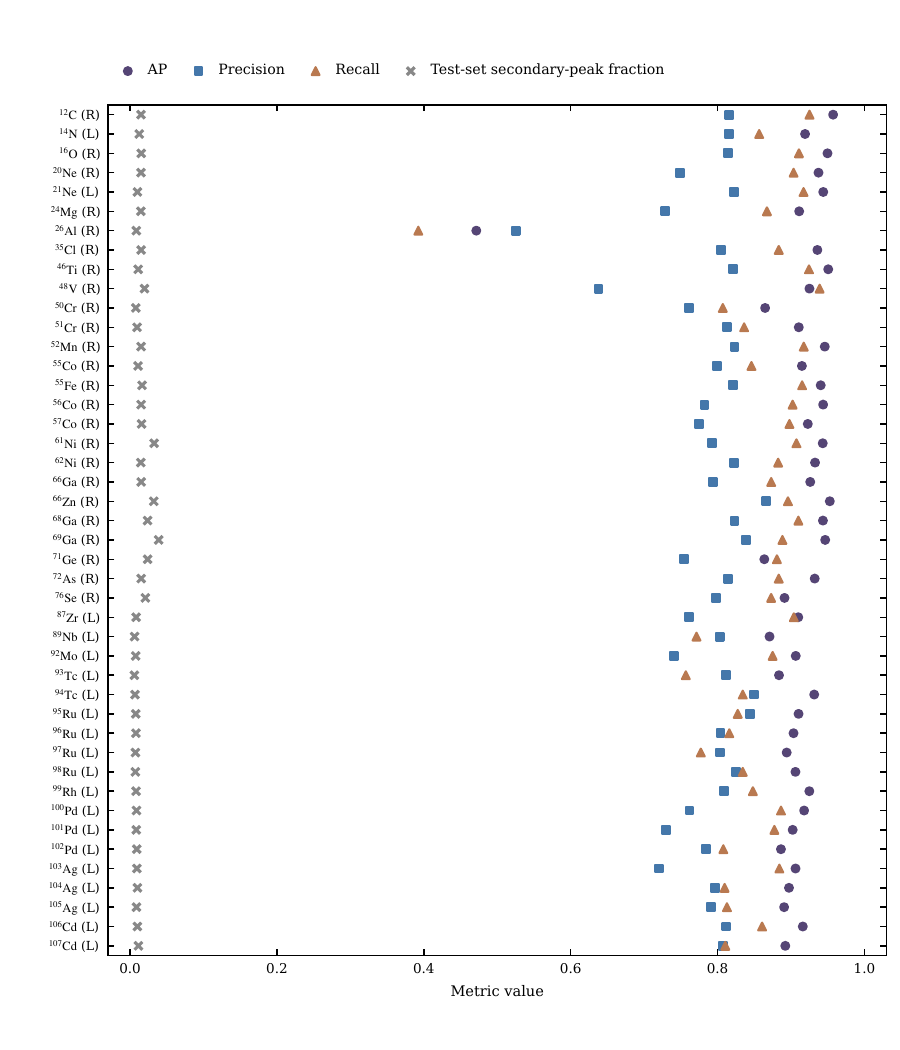}\\
\caption{\label{fig:performance}Test-set classification performance for the 44 selected isotopes, evaluated using average precision (AP), precision, and recall. Gray crosses show the fraction of secondary-peak samples in each test set. L and R indicate secondary peaks on the low- and high-abundance sides of the main peak, respectively. Of note, the classifier for \(^{26}\mathrm{Al}\) performed less well, with an AP of 0.471.}
\end{figure}

\end{minipage}
\par

\begin{figure}[H]
\centering
\includegraphics[
    width=\columnwidth,
    height=\textheight,
    keepaspectratio
]{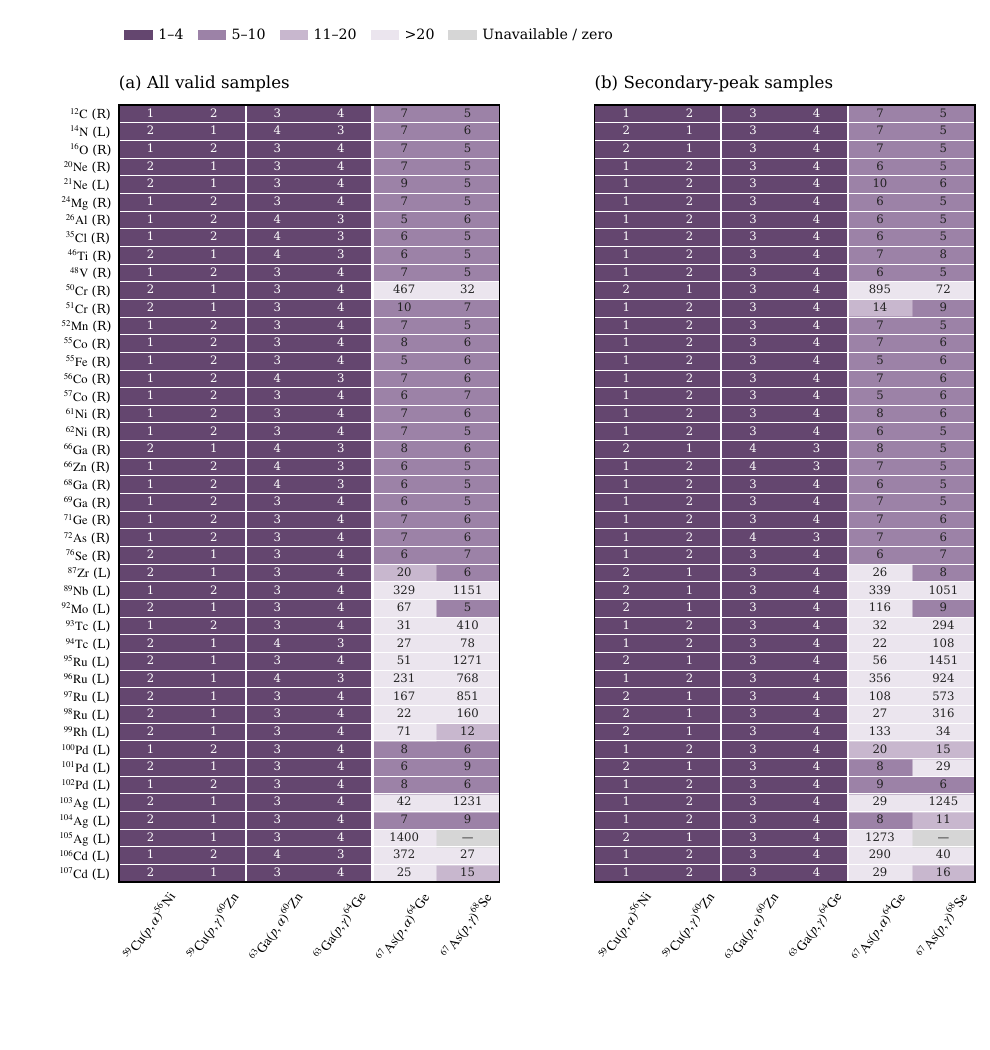}\\
\caption{\label{fig:shank rank}SHAP ranks of the \(^{59}\mathrm{Cu}\), \(^{63}\mathrm{Ga}\), and \(^{67}\mathrm{As}\) reaction pairs for the 44 selected isotopes, based on (a) all valid samples and (b) secondary-peak samples. Numbers show ranks among all input reactions, ordered by decreasing mean absolute SHAP value. Gray cells indicate unavailable ranks or zero SHAP contributions. }
\end{figure}

\clearpage

\section{Secondary-Peak Sample Fractions}
\label{app:fractions}

\begin{figure}[H]
\centering
\includegraphics[
    width=\columnwidth,
    height=0.9\textheight,
    keepaspectratio
]{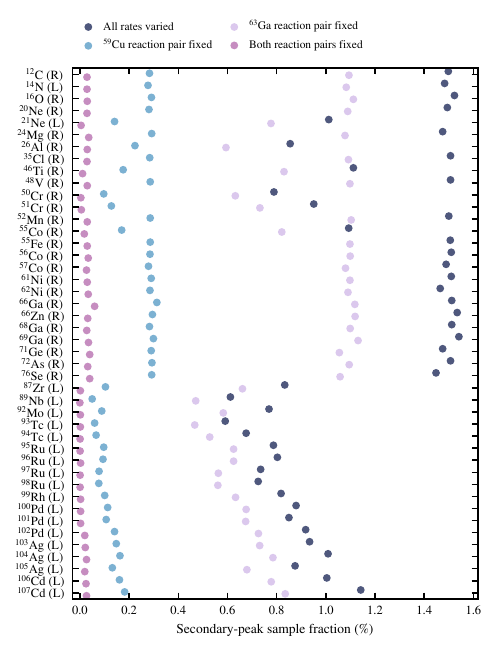}\\
\caption{\label{fig:fractions}Secondary-peak sample fractions for the 44 selected isotopes in the four sets of MC calculations. Fractions are calculated over all MC samples with finite, nonnegative final abundances. }.
\end{figure}

\begin{figure}[H]
\centering
\includegraphics[
    width=\columnwidth,
    height=0.90\textheight,
    keepaspectratio
]{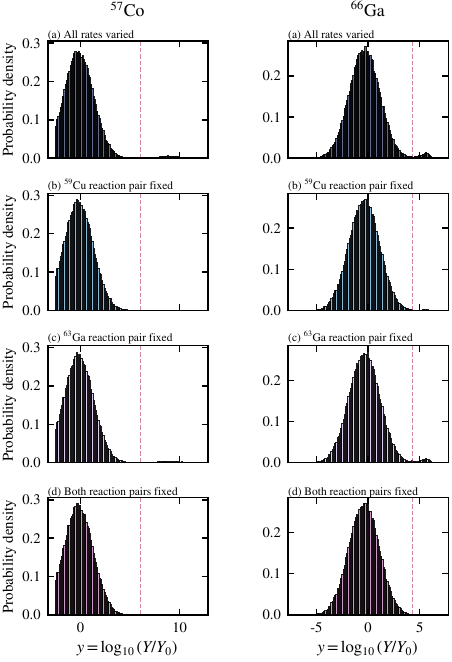}\\
\caption{\label{fig:PDF}Abundance distributions of \(^{57}\mathrm{Co}\) (left) and \(^{66}\mathrm{Ga}\) (right) for the four sets of MC calculations. Dashed lines mark the locations of minimum probability density between the two peaks in the abundance distributions from the baseline MC calculations. Samples to the right of these lines are counted as secondary-peak samples.}
\end{figure}

\clearpage

\section{Regression SHAP Analysis}
\label{app:regression_shap}

\begin{figure}[H]
\centering
\includegraphics[
    width=\columnwidth,
    height=\textheight,
    keepaspectratio
]{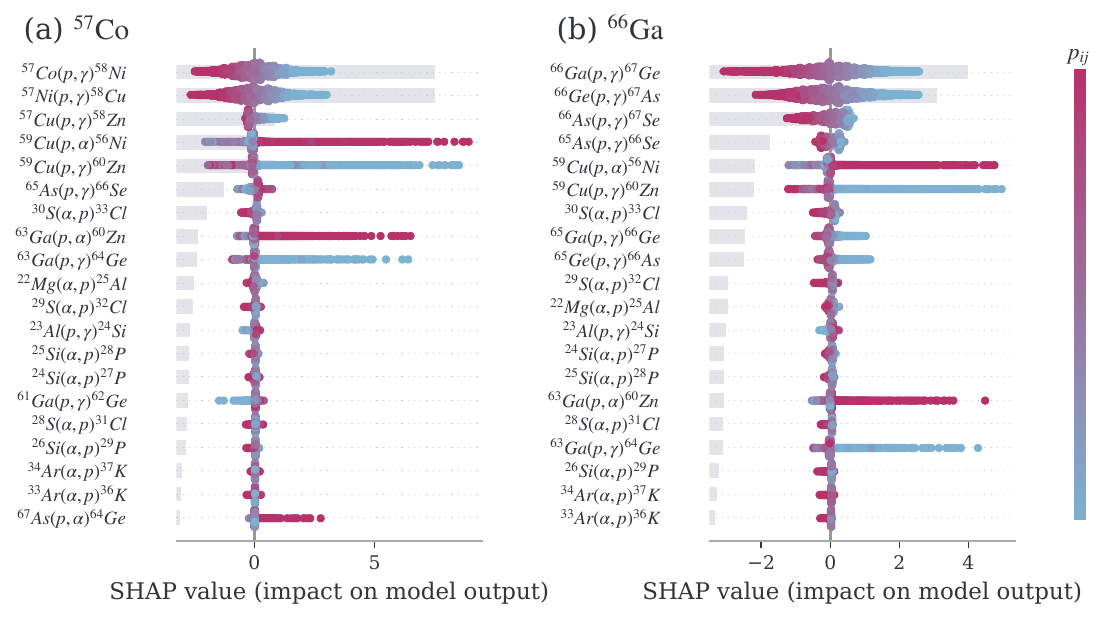}\\
\caption{\label{fig:regression_shap_co57ga66}Regression SHAP results for (a) \(^{57}\mathrm{Co}\) and (b) \(^{66}\mathrm{Ga}\) from the baseline MC calculations. The 20 reactions with the largest mean absolute SHAP values are shown in descending order, with shaded bars indicating their relative importance. Each dot represents one sample. Colors indicate the sampled reaction parameters, from low values (blue) to high values (red). Positive and negative SHAP values increase and decrease the predicted logarithmic relative abundance, respectively.}
\end{figure}

\clearpage

\bibliography{reference}{}

@article{Parikh2008,
	author = {Anuj Parikh and Jordi José and Fermín Moreno and Christian Iliadis},
	doi = {10.1086/589879},
	issn = {0067-0049},
	issue = {1},
	journal = {The Astrophysical Journal Supplement Series},
	month = {9},
	pages = {110-136},
	publisher = {American Astronomical Society},
	title = {The Effects of Variations in Nuclear Processes on Type I X‐Ray Burst Nucleosynthesis},
	volume = {178},
	url = {https://iopscience.iop.org/article/10.1086/589879 https://iopscience.iop.org/article/10.1086/589879/meta},
	year = {2008}
}

@article{Cyburt2010,
	author = {Richard H. Cyburt and A. Matthew Amthor and Ryan Ferguson and Zach Meisel and Karl Smith and Scott Warren and Alexander Heger and R. D. Hoffman and Thomas Rauscher and Alexander Sakharuk and Hendrik Schatz and F. K. Thielemann and Michael Wiescher},
	doi = {10.1088/0067-0049/189/1/240},
	issn = {0067-0049},
	issue = {1},
	journal = {The Astrophysical Journal Supplement Series},
	month = {6},
	pages = {240},
	publisher = {IOP Publishing},
	title = {THE JINA REACLIB DATABASE: ITS RECENT UPDATES AND IMPACT ON TYPE-I X-RAY BURSTS},
	volume = {189},
	url = {https://iopscience.iop.org/article/10.1088/0067-0049/189/1/240 https://iopscience.iop.org/article/10.1088/0067-0049/189/1/240/meta},
	year = {2010}
}

@article{Cyburt2016,
	author = {R. H. Cyburt and A. M. Amthor and A. Heger and E. Johnson and L. Keek and Z. Meisel and H. Schatz and K. Smith},
	doi = {10.3847/0004-637X/830/2/55},
	issn = {0004-637X},
	issue = {2},
	journal = {The Astrophysical Journal},
	month = {10},
	pages = {55},
	publisher = {IOP Publishing},
	title = {DEPENDENCE OF X-RAY BURST MODELS ON NUCLEAR REACTION RATES},
	volume = {830},
	url = {https://iopscience.iop.org/article/10.3847/0004-637X/830/2/55 https://iopscience.iop.org/article/10.3847/0004-637X/830/2/55/meta},
	year = {2016}
}

@article{Psaltis2025,
	author = {Athanasios Psaltis and Jordi José and Richard Longland and Christian Iliadis},
	doi = {10.3847/1538-4357/addfd5},
	issn = {0004-637X},
	issue = {1},
	journal = {The Astrophysical Journal},
	month = {6},
	pages = {88},
	publisher = {American Astronomical Society},
	title = {Low-metallicity nova explosions: a site for weak rp-process nucleosynthesis},
	volume = {987},
	url = {https://arxiv.org/pdf/2506.15162},
	year = {2025}
}

@article{Sallaska2013,
	author = {A. L. Sallaska and C. Iliadis and A. E. Champange and S. Goriely and S. Starrfield and F. X. Timmes},
	doi = {10.1088/0067-0049/207/1/18},
	issn = {0067-0049},
	issue = {1},
	journal = {The Astrophysical Journal Supplement Series},
	month = {7},
	pages = {18},
	publisher = {IOP Publishing},
	title = {STARLIB: A NEXT-GENERATION REACTION-RATE LIBRARY FOR NUCLEAR ASTROPHYSICS},
	volume = {207},
	url = {https://iopscience.iop.org/article/10.1088/0067-0049/207/1/18 https://iopscience.iop.org/article/10.1088/0067-0049/207/1/18/meta},
	year = {2013}
}

@article{Reichert2023,
	author = {M. Reichert and C. Winteler and O. Korobkin and A. Arcones and J. Bliss and M. Eichler and U. Frischknecht and C. Fröhlich and R. Hirschi and M. Jacobi and J. Kuske and G. Martínez-Pinedo and D. Martin and D. Mocelj and T. Rauscher and F.-K. Thielemann},
	doi = {10.3847/1538-4365/acf033},
	issn = {0067-0049},
	issue = {2},
	journal = {The Astrophysical Journal Supplement Series},
	month = {10},
	pages = {66},
	publisher = {American Astronomical Society},
	title = {The Nuclear Reaction Network WinNet},
	volume = {268},
	url = {https://ui.adsabs.harvard.edu/abs/2023ApJS..268...66R/abstract},
	year = {2023}
}

@article{Grevesse1998,
	author = {N. Grevesse and A. J. Sauval and N. Grevesse and A. J. Sauval},
	doi = {10.1023/A:1005161325181},
	issn = {0038-6308},
	issue = {1-2},
	journal = {SSRv},
	month = {5},
	pages = {161-174},
	title = {Standard Solar Composition},
	volume = {85},
	url = {https://ui.adsabs.harvard.edu/abs/1998SSRv...85..161G/abstract},
	year = {1998}
}

@article{Fields2016,
	author = {C. E. Fields and R. Farmer and I. Petermann and C. Iliadis and F. X. Timmes},
	doi = {10.3847/0004-637X/823/1/46},
	issn = {0004-637X},
	issue = {1},
	journal = {The Astrophysical Journal},
	month = {5},
	pages = {46},
	publisher = {IOP Publishing},
	title = {PROPERTIES OF CARBON–OXYGEN WHITE DWARFS FROM MONTE CARLO STELLAR MODELS},
	volume = {823},
	url = {https://iopscience.iop.org/article/10.3847/0004-637X/823/1/46 https://iopscience.iop.org/article/10.3847/0004-637X/823/1/46/meta},
	year = {2016}
}

@article{Longland2012,
	author = {R. Longland},
	doi = {10.1051/0004-6361/201220386},
	issn = {0004-6361},
	journal = {Astronomy \& Astrophysics},
	month = {12},
	pages = {A30},
	publisher = {EDP Sciences},
	title = {Recommendations for Monte Carlo nucleosynthesis sampling},
	volume = {548},
	url = {https://www.aanda.org/articles/aa/full_html/2012/12/aa20386-12/aa20386-12.html https://www.aanda.org/articles/aa/abs/2012/12/aa20386-12/aa20386-12.html},
	year = {2012}
}

@article{Koike2004,
	author = {Osamu Koike and Masa‐aki Hashimoto and Reiko Kuromizu and Shin‐ichirou Fujimoto},
	doi = {10.1086/381354},
	issn = {0004-637X},
	issue = {1},
	journal = {The Astrophysical Journal},
	month = {3},
	pages = {242-251},
	publisher = {American Astronomical Society},
	title = {Final Products of the rp‐Process on Accreting Neutron Stars},
	volume = {603},
	url = {https://iopscience.iop.org/article/10.1086/381354 https://iopscience.iop.org/article/10.1086/381354/meta},
	year = {2004}
}

@article{Winteler2012,
	author = {C. Winteler and R. Käppeli and A. Perego and A. Arcones and N. Vasset and N. Nishimura and M. Liebendörfer and F. K. Thielemann},
	doi = {10.1088/2041-8205/750/1/L22},
	issn = {2041-8205},
	issue = {1},
	journal = {The Astrophysical Journal Letters},
	month = {4},
	pages = {L22},
	publisher = {IOP Publishing},
	title = {MAGNETOROTATIONALLY DRIVEN SUPERNOVAE AS THE ORIGIN OF EARLY GALAXY r-PROCESS ELEMENTS?},
	volume = {750},
	url = {https://iopscience.iop.org/article/10.1088/2041-8205/750/1/L22 https://iopscience.iop.org/article/10.1088/2041-8205/750/1/L22/meta},
	year = {2012}
}

@article{Longland2010,
	author = {R. Longland and C. Iliadis and A. E. Champagne and J. R. Newton and C. Ugalde and A. Coc and R. Fitzgerald},
	doi = {10.1016/J.NUCLPHYSA.2010.04.008},
	issn = {0375-9474},
	issue = {1-4},
	journal = {Nuclear Physics A},
	month = {10},
	pages = {1-30},
	publisher = {North-Holland},
	title = {Charged-particle thermonuclear reaction rates: I. Monte Carlo method and statistical distributions},
	volume = {841},
	url = {https://www.sciencedirect.com/science/article/abs/pii/S0375947410004185},
	year = {2010}
}

@ARTICLE{2020SciPy-NMeth,
	author  = {Virtanen, Pauli and Gommers, Ralf and Oliphant, Travis E. and
	Haberland, Matt and Reddy, Tyler and Cournapeau, David and
	Burovski, Evgeni and Peterson, Pearu and Weckesser, Warren and
	Bright, Jonathan and {van der Walt}, St{\'e}fan J. and
	Brett, Matthew and Wilson, Joshua and Millman, K. Jarrod and
	Mayorov, Nikolay and Nelson, Andrew R. J. and Jones, Eric and
	Kern, Robert and Larson, Eric and Carey, C J and
	Polat, {\.I}lhan and Feng, Yu and Moore, Eric W. and
	{VanderPlas}, Jake and Laxalde, Denis and Perktold, Josef and
	Cimrman, Robert and Henriksen, Ian and Quintero, E. A. and
	Harris, Charles R. and Archibald, Anne M. and
	Ribeiro, Ant{\^o}nio H. and Pedregosa, Fabian and
	{van Mulbregt}, Paul and {SciPy 1.0 Contributors}},
	title   = {{{SciPy} 1.0: Fundamental Algorithms for Scientific
	Computing in Python}},
	journal = {Nature Methods},
	year    = {2020},
	volume  = {17},
	pages   = {261--272},
	adsurl  = {https://rdcu.be/b08Wh},
	doi     = {10.1038/s41592-019-0686-2},
}

@conference{jupyter,
	Title = {Jupyter Notebooks -- a publishing format for reproducible computational workflows},
	Author = {Thomas Kluyver and Benjamin Ragan-Kelley and Fernando P{\'e}rez and Brian Granger and Matthias Bussonnier and Jonathan Frederic and Kyle Kelley and Jessica Hamrick and Jason Grout and Sylvain Corlay and Paul Ivanov and Dami{\'a}n Avila and Safia Abdalla and Carol Willing},
	Booktitle = {Positioning and Power in Academic Publishing: Players, Agents and Agendas},
	Editor = {F. Loizides and B. Schmidt},
	Organization = {IOS Press},
	Pages = {87 - 90},
	Year = {2016}
}

@Article{matplotlib,
	Author    = {Hunter, J. D.},
	Title     = {Matplotlib: A 2D graphics environment},
	Journal   = {Computing in Science \& Engineering},
	Volume    = {9},
	Number    = {3},
	Pages     = {90--95},
	publisher = {IEEE COMPUTER SOC},
	doi       = {10.1109/MCSE.2007.55},
	year      = 2007
}

@Article{numpy,
	title         = {Array programming with {NumPy}},
	author        = {Charles R. Harris and K. Jarrod Millman and St{\'{e}}fan J.
	van der Walt and Ralf Gommers and Pauli Virtanen and David
	Cournapeau and Eric Wieser and Julian Taylor and Sebastian
	Berg and Nathaniel J. Smith and Robert Kern and Matti Picus
	and Stephan Hoyer and Marten H. van Kerkwijk and Matthew
	Brett and Allan Haldane and Jaime Fern{\'{a}}ndez del
	R{\'{i}}o and Mark Wiebe and Pearu Peterson and Pierre
	G{\'{e}}rard-Marchant and Kevin Sheppard and Tyler Reddy and
	Warren Weckesser and Hameer Abbasi and Christoph Gohlke and
	Travis E. Oliphant},
	year          = {2020},
	month         = sep,
	journal       = {\nat},
	volume        = {585},
	number        = {7825},
	pages         = {357--362},
	doi           = {10.1038/s41586-020-2649-2},
	publisher     = {Springer Science and Business Media {LLC}},
	url           = {https://doi.org/10.1038/s41586-020-2649-2}
}

@article{scikit-learn,
	title={Scikit-learn: Machine Learning in {P}ython},
	author={Pedregosa, F. and Varoquaux, G. and Gramfort, A. and Michel, V.
	and Thirion, B. and Grisel, O. and Blondel, M. and Prettenhofer, P.
	and Weiss, R. and Dubourg, V. and Vanderplas, J. and Passos, A. and
	Cournapeau, D. and Brucher, M. and Perrot, M. and Duchesnay, E.},
	journal={Journal of Machine Learning Research},
	volume={12},
	pages={2825--2830},
	year={2011}
}

@Article{ipython,
	Author    = {P\'erez, Fernando and Granger, Brian E.},
	Title     = {{IP}ython: a System for Interactive Scientific Computing},
	Journal   = {Computing in Science and Engineering},
	Volume    = {9},
	Number    = {3},
	Pages     = {21--29},
	month     = may,
	year      = 2007,
	url       = "https://ipython.org",
	ISSN      = "1521-9615",
	doi       = {10.1109/MCSE.2007.53},
	publisher = {IEEE Computer Society},
}

@article{Waskom2021,
	doi = {10.21105/joss.03021},
	url = {https://doi.org/10.21105/joss.03021},
	year = {2021},
	publisher = {The Open Journal},
	volume = {6},
	number = {60},
	pages = {3021},
	author = {Michael L. Waskom},
	title = {seaborn: statistical data visualization},
	journal = {Journal of Open Source Software}
}

@article{Lam2016,
   author = {Y. H. Lam and J. J. He and A. Parikh and H. Schatz and B. A. Brown and M. Wang and B. Guo and Y. H. Zhang and X. H. Zhou and H. S. Xu},
   doi = {10.3847/0004-637X/818/1/78},
   issn = {0004-637X},
   issue = {1},
   journal = {The Astrophysical Journal},
   month = {2},
   pages = {78},
   publisher = {IOP Publishing},
   title = {REACTION RATES OF 64Ge(p,γ)65As AND 65As(p,γ)66Se AND THE EXTENT OF NUCLEOSYNTHESIS IN TYPE I X-RAY BURSTS},
   volume = {818},
   url = {https://iopscience.iop.org/article/10.3847/0004-637X/818/1/78 https://iopscience.iop.org/article/10.3847/0004-637X/818/1/78/meta},
   year = {2016}
}

@article{Lam2022b,
   author = {Yi Hua Lam and Zi Xin Liu and Alexander Heger and Ning Lu and Adam Michael Jacobs and Zac Johnston},
   doi = {10.3847/1538-4357/AC4D8B},
   issn = {0004-637X},
   issue = {1},
   journal = {The Astrophysical Journal},
   month = {4},
   pages = {72},
   publisher = {IOP Publishing},
   title = {Impact of the New 65As(p,γ)66Se Reaction Rate on the Two-proton Sequential Capture of 64Ge, Weak GeAs Cycles, and Type I X-Ray Bursts Such as the Clocked Burster GS 1826−24},
   volume = {929},
   url = {https://iopscience.iop.org/article/10.3847/1538-4357/ac4d8b https://iopscience.iop.org/article/10.3847/1538-4357/ac4d8b/meta},
   year = {2022}
}

@article{Lam2022c,
   author = {Yi Hua Lam and  Ning Lu and Alexander Heger and Adam Michael Jacobs and Nadezda A. Smirnova and Teresa Kurtukian Nieto and Zac Johnston and Shigeru Kubono},
   doi = {10.3847/1538-4357/AC4D89},
   issn = {0004-637X},
   issue = {1},
   journal = {The Astrophysical Journal},
   month = {4},
   pages = {73},
   publisher = {IOP Publishing},
   title = {The Regulated NiCu Cycles with the New 57Cu(p,γ)58Zn Reaction Rate and Its Influence on Type I X-Ray Bursts: the GS 1826–24 Clocked Burster},
   volume = {929},
   url = {https://iopscience.iop.org/article/10.3847/1538-4357/ac4d89 https://iopscience.iop.org/article/10.3847/1538-4357/ac4d89/meta},
   year = {2022}
}

@article{Wang2026,
author = {Q. Wang and E. T. Li and Zhihong Li and Youbao Wang and Bing Guo and Yunju Li and Jun Su and Shipeng Hu and Yinwen Guan and Dong Xiang and Yu Liu and Lei Yang and Weiping Liu},
title = {Impact of Nuclear Reaction Rate Uncertainties on Type I X-ray Burst Nucleosynthesis: A Monte Carlo Study},
journal = {The Astrophysical Journal},
year = {2026},
doi = {10.3847/1538-4357/ae4fb7},
note = {Accepted for publication}
}

@article{Lundberg2019,
author  = {Lundberg, Scott M. and Erion, Gabriel G. and Lee, Su-In},
title   = {Consistent Individualized Feature Attribution for Tree Ensembles},
journal = {arXiv e-prints},
year    = {2019},
eprint  = {1802.03888},
archivePrefix = {arXiv}
}

@article{Lundberg2020,
author  = {Lundberg, Scott M. and Erion, Gabriel G. and Chen, Hugh and DeGrave, Alex and Prutkin, Jordan M. and Nair, Balasubramanian and Katz, Ronit and Himmelfarb, Jonathan and Bansal, Nisha and Lee, Su-In},
title   = {From local explanations to global understanding with explainable AI for trees},
journal = {Nature Machine Intelligence},
volume  = {2},
pages   = {56--67},
year    = {2020},
doi     = {10.1038/s42256-019-0138-9}
}

@inproceedings{Chen2016,
author = {Chen, Tianqi and Guestrin, Carlos},
title = {{XGBoost}: A Scalable Tree Boosting System},
booktitle = {Proceedings of the 22nd ACM SIGKDD International Conference on Knowledge Discovery and Data Mining},
series = {KDD '16},
year = {2016},
pages = {785--794},
publisher = {Association for Computing Machinery},
address = {New York, NY, USA},
doi = {10.1145/2939672.2939785},
url = {https://doi.org/10.1145/2939672.2939785}
}

@article{Saito2015,
author = {Saito, Takaya and Rehmsmeier, Marc},
title = {The Precision-Recall Plot Is More Informative than the {ROC} Plot When Evaluating Binary Classifiers on Imbalanced Datasets},
journal = {PLOS ONE},
year = {2015},
volume = {10},
number = {3},
pages = {e0118432},
doi = {10.1371/journal.pone.0118432},
url = {https://doi.org/10.1371/journal.pone.0118432}
}

@article{Merz2021,
  author  = {Merz, Grant and Meisel, Zach},
  title   = {Urca nuclide production in {Type-I} {X-ray} bursts and implications for nuclear physics studies},
  journal = {Monthly Notices of the Royal Astronomical Society},
  volume  = {500},
  number  = {3},
  pages   = {2958--2968},
  year    = {2021},
  doi     = {10.1093/mnras/staa3414}
}

@article{Bhathi2026,
  author = {Bhathi, N. and Randhawa, J. S. and Kanungo, R. and Refsgaard, J. and Alcorta, M. and Ahn, T. and Andreoiu, C. and Bardayan, D. and Bhattacharjee, S. S. and Davids, B. and Christian, G. and Chen, A. A. and Coleman, R. and Garrett, P. E. and Grinyer, G. F. and {Gyabeng Fuakye}, E. and Hackman, G. and Jain, R. and Kapoor, K. and Kr{\"u}cken, R. and Laffoley, A. and Lennarz, A. and Liang, J. and Meisel, Z. and Psaltis, A. and Radich, A. and Rocchini, M. and Rojo, J. S. and Saei, N. and Saxena, M. and Singh, M. and Svensson, C. E. and Subramaniam, P. and Talebitaher, A. and Upadhyayula, S. and Waterfield, C. and Williams, J. and Williams, M. and Zubair, M. A.},
  title = {Direct Measurement of {$^{59}$Cu(p,$\alpha$)$^{56}$Ni} Precludes a Strong {NiCu} Cycle in {Type-I} {X-Ray} Bursts},
  journal = {The Astrophysical Journal},
  volume = {999}, number = {1}, pages = {8}, year = {2026},
  doi = {10.3847/1538-4357/ae3de6}
}

@article{LopezSaavedra2026Short,
  author = {Lopez-Saavedra, E. and Avila, M. L. and Ong, W.-J. and Mohr, P. and Psaltis, A. and Ahn, S. and Arora, H. and Balliet, L. and Bhatt, K. and Cha, S. M. and Chipps, K. A. and Dopfer, J. and Tolstukhin, I. A. and Jain, R. and Kim, M. J. and Kolos, K. and Montes, F. and Neto, D. and Pain, S. D. and Pereira, J. and Randhawa, J. S. and Sun, L. J. and Ugalde, C. and Wagner, L.},
  title = {Direct Measurement of the {$^{59}$Cu$(p,\alpha)^{56}$Ni} Excitation Function to Constrain the {Ni--Cu} Cycle Strength and Its Impact on Explosive Nucleosynthesis},
  journal = {arXiv e-prints}, pages = {arXiv:2601.22262}, year = {2026},
  eprint = {2601.22262}, archivePrefix = {arXiv}, primaryClass = {nucl-ex},
  doi = {10.48550/arXiv.2601.22262}
}

@article{LopezSaavedra2026Long,
  author = {Lopez-Saavedra, E. and Avila, M. L. and Ong, W.-J. and Mohr, P. and Ahn, S. and Arora, H. and Balliet, L. and Bhatt, K. and Cha, S. M. and Chipps, K. A. and Dopfer, J. and Tolstukhin, I. A. and Jain, R. and Kim, M. J. and Kolos, K. and Montes, F. and Neto, D. and Pain, S. D. and Pereira, J. and Randhawa, J. S. and Sun, L. J. and Ugalde, C. and Wagner, L.},
  title = {Detailed Study of the {$^{59}$Cu(p,$\alpha$)$^{56}$Ni} Reaction and Constraints on Its Astrophysical Reaction Rate},
  journal = {arXiv e-prints}, pages = {arXiv:2601.22279}, year = {2026},
  eprint = {2601.22279}, archivePrefix = {arXiv}, primaryClass = {nucl-ex},
  doi = {10.48550/arXiv.2601.22279}
}

@article{Xu2026,
    author  = {Xu, Yi and Schatz, Hendrik and Lau, Rita and Meisel, Zach and Mohr, Peter},
    title   = {Nuclear Physics of X-ray Bursts},
    journal = {arXiv preprint arXiv:2603.05356},
    year    = {2026},
    eprint  = {2603.05356},
    archivePrefix = {arXiv},
    primaryClass  = {astro-ph.HE},
    doi     = {10.48550/arXiv.2603.05356},
    url     = {https://arxiv.org/abs/2603.05356v1},
    note    = {Accepted for publication in Physics Reports}
}
\bibliographystyle{aasjournalv7}
\end{document}